\documentstyle[preprint,aps]{revtex}
\begin{document}
\preprint{IP/BBSR/94-12,hep-th/yymmnn}
\title{Exact self dual vortices in $d=3$ Majumdar-Papapetrou
dilaton gravity}
\author{C.S.Aulakh}
\address{Institute of Physics, Bhubaneshwar, Orissa 751005, India}
\maketitle
\begin{abstract}
We show that self-dual Nielsen Olesen (NO)
 vortices in $3$ dimensions give rise
to a class of exact solutions when coupled to Einstein Maxwell Dilaton gravity
obeying the Majumdar-Papapetrou(MP)
 relation between gravitational and Maxwell
couplings , provided certain Chern-Simons type interactions are present.
The metric may be solved for explicitly in terms of the NO vortex function
and is asymptotic to Euclidean space with signature (-1,-1,-1).
The MP electric field is long range but, strictly speaking, the charge of the
vortices is zero since the field dies off as  $O(1/r (ln r)^2)$.
The total ADM energy integral of such vortices is {\it{zero}}.
 These peculiarities are due
to the nature of the two dimensional Greens function.
\end{abstract}
\pacs{PACS numbers:  }

\def\pdp{\psi^{\dagger}\psi}
\def\a{\alpha}
\def\b{\beta}
\def\g{\gamma}
\def\d{\delta}
\def\e{\epsilon}
\def\t{\theta}
\def\k{\kappa}
\def\l{\lambda}
\def\om{\omega}
\def\p{\pi}
\def\m{\mu}
\def\n{\nu}
\def\r{\rho}
\def\s{\sigma}
\def\rt{${\tilde r}$}
\def\lnrt{\longrightarrow}
\def\prt{\partial}
\def\prmu{\partial_{\m}}
\def\prlam{\partial_{\l}}

\narrowtext
 In an astonishing recent paper \cite{gibbons} it was shown that the self dual
``instantonic'' solitons of $4+1$ dimensional Yang-Mills (YM) theory and the
self dual BPS \cite{bogo} monopoles of $3+1$ dimensional
Yang-Mills-Higgs theory give rise to a class of exact stable static solutions
even after they are coupled to Einstein-Maxwell(EM)  gravity at the Majumdar-
Papapetrou(MP) point\cite{MP} (i.e the Maxwell coupling is fixed by the
gravitational
coupling so that the static Newton and Couloumb potentials are equal
in magnitude).
Important
additional requirements are the inclusion of certain non-minimal terms of
form $\e^{MNLPQ} A_M F_{NL}^A F_{PQ}^A$ in $4+1$ dimensions (and  its
dimensional reduction in $3+1$ dimensions) and coupling to a dilaton in $3+1$
dimensions. Here  $A_M,A_M^A$ are the Electromagnetic and YM gauge potentials
and $F_{MN}^A$ is the YM field strength . These terms are responsible for
the soliton aquiring a topological electric charge coupled to the Maxwell
field by a mechanism proposed similar to one proposed for $4+1$ Yang Mills
Chern Simons(YMCS) theory solitons in \cite{syn,cts,tsm}.
This  electric charge acts as a Bogomol'nyi bound under the ADM \cite{adm}
mass of field configurations in the theory.
 The self dual solitonic solutions of
Ref.[1] saturate this bound insuch a way as to imply the existence of a
Killing spinor with respect to a certain Einstein Maxwell covariant
derivative. Since the bosonic action used is a subset of the $d=5, N=2$
supergravity \cite{gunyadin} it follows that the solutions preserve one
 of the two supersymmetries thus providing yet another example of the
supersymmetry associated with self dual solutions of field equations
\cite{sdsusy}. The result of Ref.[1] allowed us to immedeately confirm
\cite{tep} the  conjecture \cite{cts,tsm} concerning the gravitational
stabilization of instantonic configurations in $SU(N), N\geq 3$ YMCS theory.
 The elegance and naturalness of the arguments of Ref.[1] lead one to expect
 that generalizations to other dimensions and types of solitons (strings,
vortices etc.) should exist. In this letter we obtain solutions
in $3$ dimensions analogous  to those  of Ref.[1] in four and five
 dimensions. However the peculiarities of the spherically symmetric
Greens function in two spatial dimensions (i.e the logarithm)
lead to some peculiar properties for the solutions we generate. The asymptotic
spacetime is flat but Euclidean (with no deficit angle ). Correspondingly, the
total ADM energy integral yields {\it{zero}} and the MP
 electric field, though long range, dies off
marginally faster (by a factor of $1/(log r)^2$ ) than that of a charged
object. Thus  the expected topological charge is also zero. Nevertheless
our solutions appear sufficently similar to those of Ref.[1] to keep alive
the  expectation of generalizations to other models where flatspace
self-dual solutions exist .

The action we shall treat is the sum of three pieces $S_{gr}, S_{SD}$
 and $S_{CS}$ :

\begin{equation}
S_{gr}= -(16 \p G)^{-1} \int d^3 x E ( R +{e^{2 b \s }} F^2 +
2 (\prt \s )^2)
\end{equation}
\begin{equation}
S_{SD} =  -\int d^3 x{}E{} (\frac{1}{4 g^2}{e^{2 b_a \s }} f^2 +
{e^{2 b_{\psi} \s }} \mid D\psi\mid^2 +\frac{ g^2}{2} {e^{2 b_u \s }}
(\psi^{\dagger}\psi -v^2)^2)
\end{equation}
\begin{equation}
S_{CS} =\int d^3 x\quad \epsilon^{\m \n \l} A_{\m}
({\k}_{1} {\prt}_{\n} a_{\l} +{\k}_2 {\prt}_{\n} J_{\l})
\end{equation}

Here $F,f$ are field strengths of the MP ($A_{\m}$) and NO ($a_{\m}$)
Abelian gauge potentials, $\psi$
the charged scalar field of the NO model, $\s$ the dilaton field and $J_{\l}
={\frac{1}{2i}}\psi^{\dagger}{\stackrel{\leftrightarrow}{D_{\l}}}\psi,
( D_{\l}=\partial_{\l}\psi -i a_{\l}\psi)$ the NO current.
Notice that the MP coupling has been equated to the gravitational one :
$g_e^2 =4 \p G$. The dilaton couplings $b,b_a,b_{\psi},b_u$
will be chosen in the course of the calculation to allow an exact solution
of the full theory given a flat space self dual NO vortex solution. Note,
 however, that the different terms in the action do {\it{not}}
scale in the same way  under  the rescaling
$\s \rightarrow \s + const$ for any choice of
scaling weights for the other fields unless one also scales couplings.
The other possible Chern-Simons terms
can be added to $S_{CS}$ without affecting our conclusions .
Our conventions for gravitational quantities are those of \cite{weinberg}
E is the determinant of the {\it{dreibein}} while $\e^{\m\n\l}$ is the
3-d antisymmetric tensor density ($\e^{012}=1$).

To proceed we note that the MP form of the metric in $D+1$ dimensions can
be written in a form equivalent to that of \cite{myers} as ($i=1,2$) :

\begin{equation}
ds^2=-\frac{1}{B^{D-2}} dt^2 + B(x^1,x^2) dx^i dx^i
\end{equation}

While this form of the metric can also be derived from considerations
of the existence of a Killing spinor \cite{gibbons}, one can view this ansatz
simply as dictated by the need to ensure that the off diagonal terms of the
Einstein tensor are automatically zero given that the spatial metric is
conformal to the Euclidean one. In $2+1$ dimensions we therefore take
the metric to be $Diag(-1,B,B)$. We also write $B=e^{2\phi}$

 The field equations are :
\begin{equation}
{\tilde G}_{\m\n} =G_{\m\n} + 8\p G T_{\m\n}(A,\s)=-8\p
G T^{SD}_{\m\n}(a_{\m},\psi)
\end{equation}
\begin{equation}
{\prt}_{\m} (Ee^{2b\s} F^{\m\n})=-4\p G{\e^{\n\m\l}}\prt_{\m}(\k_1 a_{\l} +
\k_2 J_{\l})
\end{equation}

\begin{eqnarray}
\prt_{\m}(E g^{\m\n}\prt_{\n}\s) - (b/2){e^{2b\s}}E F^2
&=&8\p G E((b_a/4g^2) {e^{2b_a\s}} f^2 \nonumber\\
&+& b_{\psi}{e^{2 b_{\psi} \s }} \mid D\psi\mid^2 +(g^2/2) b_u
 {e^{2 b_u \s }}  (\psi^{\dagger}\psi -v^2)^2)
\end{eqnarray}

\begin{equation}
\prt_{\m}({e^{2b_a\s}} E f^{\m\n}) =-2 g^2 E {e^{2b_{\psi}\s}} J^{\n} -
g^2 \e^{\n\m\l}(\prt_{\m}A_{\l} (\k_1 -\k_2 \psi^{\dagger}\psi) )
\end{equation}
\begin{equation}
D_{\m}({e^{2b_{\psi}\s}} E D^{\m}\psi) ={e^{2b_u\s}} E \frac{\prt U}
{\prt{\psi^{\dagger}}} +
i\k_2 \e^{\n\m\l}\prt_{\n}A_{\m} D_{\l}\psi
\end{equation}
 where $U=(g^2/2)(\psi^{\dagger}\psi -v^2)^2$ is the potential.

When the metric is flat and $\s=0,\k_i=0$ the model reduces to the selfdual
Abelian Higgs  model whose field equations are solved and the energy
minimized provided the fields are static, $a_0=0$,
 and the Bogomol'nyi equations:
\begin{equation}
(D_i \mp i\e_{ij} D_j)\psi =0
\end{equation}
\begin{equation}
 f_{ij} =\pm\e_{ij} g^2 (\psi^{\dagger}\psi -v^2)
\end{equation}
 are satisfied. These first order equations may be decoupled to
yield the well known vortex equation \cite{jaffe} :

\begin{equation}
\prt^2 ln \frac{\psi^{\dagger}\psi}{v^2} =g^2 (\psi^{\dagger}\psi -v^2)
+ 4\p \sum_k\mid n_k\mid \d^{(2)}({\vec r}-{\vec r}_k)
\end{equation}
where $\{n_k,{\vec r_k}\}$ are the winding numbers and locations
(positions of the zeros of $\psi$) of an ensemble of (anti)self dual vortices.

We now show that with a certain choice of dilaton weights every solution of
the above self duality equations gives rise to an {\it{exact, explicit}}
 solution of  the field equations of the full theory. We impose the relation :

\begin{equation}
{e^{2b_{\psi}\s}}={e^{2b_a\s}} B^{-1} = B {e^{2b_u\s}}
\end{equation}
which will be satisfied for a certain choice of weights provided $\s=\d \phi$.
Then it is easy to see, using the assumed flatspace self duality of the NO
fields, that provided
\begin{eqnarray}
A_0&=& \mp\frac{E {e^{2b_{\psi}\s}}}{\k_2 B}\nonumber \\
A_i&=&0
\end{eqnarray}
and ${\k_1}/{\k_2}=v^2$
the field equations for the fields $a_{\m},\psi$ are satisfied. Furthermore
it is easy to check that all spatial components of the
 Einstein tensor and the stress tensor of the matter sector vanish . Hence
the spatial components of the Einstein equations are satisfied provided
the spatial stress tensor of the fields $A_{\m},\s$ vanishes  which requires
that
\begin{equation}
{e^{2b_{\psi}\s}} = \frac{s {\bar s} \k_2 {e^{-b\s}}}{b }
\end{equation}
where $s=\pm$ is an independent sign and the reason for being careful
 about putting $E/B=sign(B)={\bar s}$ (although we took $B=e^{2\phi}$!) will
become clear {\it{a posteriori}}. Thus one immedeately has
$b_{\psi}=-b/2, \k_2={\bar s} s b$.

Since we are looking for a smooth soliton solution with a dispersed energy
density we do not expect any zeros of the metric to occur . We therefore
 assume that the sign of B is constant . This will be justified
{\it{a posteriori}} and can, in any case, be relaxed (if one wishes to study
superpositions of vortices and black holes etc.) by taking the equation we
 derive to hold piece wise in every region labelled by a given sign of B
(since our equation is independent of the sign of B).
One finds that three remaining nontrivial field equations
(i.e those for $G_{00}, A_0, \s$) reduce to the {\it{single}} flat space
  equation :
\begin{eqnarray}
\prt^2 e^{2\phi}&=&\pm
(16 \p G)\e_{ij}(\frac{v^2 f_{ij}}{2} +\prt_i J_j)
\nonumber \\ &\equiv &(16 \p G) v^2(g^2 (\psi^{\dagger}\psi -v^2) -
\frac{1}{2}\prt^2(\frac{\psi^{\dagger}\psi}{v^2}))
\end{eqnarray}
{\it{if and only if}} one identifies $\s=\phi, b=2$ so that $b_{\psi}=-1,
b_a=0, b_U=-2$. Now, using the vortex equation, it immedeately follows that
the regular solution of the above Poisson equation is:

\begin{equation}
e^{2\phi} =-C + \m(log(\frac{\psi^{\dagger}\psi}{v^2})
- \frac{\psi^{\dagger}\psi}{2 v^2} -
2\sum_k \mid n_k\mid log(v^2 \mid\vec r- \vec r_k\mid) )
\end{equation}
where $C$ is an arbitrary constant and $\m=16 \p G v^2 >0$.
 Thus it follows that as
$r\rightarrow \infty,B\rightarrow -2\m \mid N\mid log(r); (\sum_k n_k=N)$!!
With suitable C , B is negative everywhere since $\psi^{\dagger}\psi$
grows from $0$ at the locations of the vortices to  $v^2$ at $\infty$. A
sensible choice of asymptotic coordinates is thus
$\tilde r= r {\sqrt{2 \m \mid N\mid log r}} $
which , to leading order in $1/log{}r$
converts the metric to a flat metric with signature $(-1,-1,-1)$ !. Note that
apart from implying that the dilaton field has a constant  imaginary part :
$Im(\s)=i\p/2$ (which is consistent with the reality of the action etc), there
is no other {\it{outre}} consequence of our solution.

The ADM ``energy'' (since our metric has signature $(-1,-1,-1)$ there
is really no energy in the usual sense) integral gives

\begin{equation}
{\cal{E}}=- \int d^2 x T^0_0 \mid B\mid \equiv -
\frac{\bar s}{16 \p G} \int d^2 x \prt^2\phi
\end{equation}
and thus vanishes given the smoothness and regularity of our solution.

Finally, the asymptotic electric field in the new coordinates is
\begin{equation}
F^{{\tilde r}0} =\mp\frac{s}{4\m \mid N\mid {\tilde r} (log {\tilde r})^2}
\end{equation}
so that its surface integral dies logarithmically.

The pathologies of 3 dimensions and the special couplings
of our `dilaton' are  responsible for the peculiarities of our solution.
Nevertheless as an addition to the library of exact solutions of
gravitaionally coupled theories it merits further investigation and motivates
the search for further examples of the mechanism of Ref.[1] which involves
as an essential part the aquisition of {\it{topological electric fields}} by
magnetic configurations in the presence of CS type couplings
\cite{syn,cts,tsm,gibbons} .
We have also found flatpace (Minkowski) versions of these results (and
of those for d=4+1 in Ref.[10]) by choosing suitable dilaton couplings.

\acknowledgements

I am grateful to Avinash Khare for useful and stimulating discussions.

\eject

\begin{references}
\bibitem{gibbons} G.W.Gibbons, D.Kastor, L.A.J.London,
P.K.Townsend and\hfil\break   J.Traschen, hep-th/9310118.
\bibitem{bogo} E.B.Bogomol'nyi, Sov. J. Nucl. Phys. {\bf{24}},449(1976);
M.K. Prasad and C.M. Sommerfield, Phys. Rev. Lett. {\bf{35}},760(1975).
\bibitem{MP} S.D.Majumdar, Phys.Rev.{\bf{72}},390(1947); A.Papapetrou,
Proc. R. Irish Acad. {\bf{A51}},191(1947).
\bibitem{syn} C.S.Aulakh, Mod. Phys. Lett {\bf{A7}},2119(1992).
\bibitem{cts} C.S.Aulakh, Mod. Phys. Lett. {\bf{A7}},2469(1992).
\bibitem{tsm} C.S.Aulakh and V.Soni, Int. J.Mod.Phys. {\bf {A8}},1653(1993).
\bibitem{adm} R.Arnowitt, S.Deser and C.Misner,
 Phys. Rev. {\bf{117}},1595(1960); Phys. Rev. {\bf{118}},1100(1960);
 Phys. Rev. {\bf{122}},997(1961).
\bibitem{gunyadin} M.G{\"u}nyadin, G.Sierra, and P.K.Townsend, Nucl. Phys.
 {\bf{B242}}, 244(1985); {\it{ibid}} {\bf{B253}}, 573(1985).
\bibitem{sdsusy} E. Witten and D. Olive, Phys. Lett. {\bf{B78}},97 (1978);
Z.Hlousek and D.Spector, Nucl. Phys. {\bf{B370}},143 (1992);
Phys. Lett. {\bf{B283}},75(1992).
\bibitem{tep}  {\it{Topological electropoles in
$4+1$ dimensional EYMCS theory}}
, C.S.Aulakh and V.Soni, IP/BBSR/94-11, hep-th 9403085.
\bibitem{weinberg} S.Weinberg, {\it{Gravitation and Cosmology}},
 (Wiley, New York, 1972).
\bibitem{myers} R.C.Myers, Phys. Rev. {\bf{D35}},455(1987).
\bibitem{jaffe} A.Jaffe and C.Taubes {\it{Vortices and Monopoles}}
(Birkhauser, Boston ,1980).


\end{references}
\end{document}